\documentstyle[twocolumn,aps,epsf]{revtex}

\newcommand{\Av}{\mbox{{\bf A}}}

\newcommand{\Rv}{\mbox{{\bf R}}}

\newcommand{\lv}{\mbox{{\bf l}}}

\input{epsf}
\begin{document}
\draft
\title{Upper critical field for anisotropic superconductivity. 
A tight--binding approach.}
\author{Maciej M. Ma\'ska and Marcin Mierzejewski\thanks{marcin@phys.us.edu.pl}}
\address{
Department of Theoretical Physics, Institute of Physics, 
University of Silesia, 40-007 Katowice,
Poland}
\maketitle
\begin{abstract}
We study the problem of the upper critical field ($H_{c2}$) for 
tight--binding electrons in a two--dimensional lattice. The external
magnetic field is introduced into the model Hamiltonian both via the
Peierls substitution and the Zeeman term. Carrying out calculations 
for finite systems we analyze the influence of the external field in 
the commensurable and incommensurable case on an equal footing.  
The upper critical field has been discussed for intrasite 
as well as anisotropic intersite pairing that, in the absence of 
magnetic field, has a $d_{x^2-y^2}$ symmetry. 
A comparison of $H_{c2}$ determined for different
symmetries shows that the on--site pairing is more affected
by the external field i.e., the critical temperature for the on--site
pairing decreases with the increase of the magnetic field faster 
than in the anisotropic case. Moreover, we have shown that the 
tight--binding form of the Bloch energy can lead to the upward 
curvature of $H_{c2}$, provided that the Fermi level is close enough
to the 
van Hove singularity.
\end{abstract}
\section{Introduction}

One of many striking  properties of high--temperature superconductors
is related to the field--induced transition from superconducting 
to normal state. Magnetic properties of high--$T_c$ compounds    
give rise to both quantitative and qualitative differences with respect
to the conventional superconductors. The systems under consideration
are characterized by extremely high values of the upper critical
 and its unusual temperature dependence.
For optimally doped
samples experimental investigation of the critical field is limited only
to temperatures close to $T_c$ \cite{osofsky}, whereas at lower temperatures 
the magnitude of $H_{c2}$ is far beyond the reach of laboratory magnetic fields.
The measurements carried out in a wide range of temperature 
for underdoped superconductors clearly indicate the positive curvature
of $H_{c2}(T)$ even at genuinely low temperatures \cite{osofsky,mackenzie,Iye,Joy}.
Theoretical approaches do not provide a unique, complete description
of these phenomena. The most of unconventional properties of high-temperature 
superconductors, like narrow quasiparticle bands, lifetime effects 
of states close to the Fermi level and linear temperature dependence of
the normal--state resistivity are usually attributed to strong
Coulomb correlations. However, upward curvature of the upper critical field
is observed also in overdoped compounds, where the temperature dependence
of resistivity changes gradually from linear to quadratic behavior \cite{kubo,suzuki}. 
This feature suggests
that the positive curvature of $H_{c2}(T)$ could originate from  e.g.,
symmetry of the superconducting order parameter or details of the density
of states and may  be explained without
a sophisticated treatment of the most difficult problem that is related to the
presence of strong electronic correlations. 
    
It is believed that the symmetry of the superconducting state can be close related to 
the pairing mechanism. There is a lot of node--sensitive experiments, 
based on the angle resolved photoemission spectroscopy \cite{arpes}, London 
penetration depth \cite{london},
NMR \cite{nmr} and quasiparticle tunneling \cite{tunel},
which indicate that the energy gap is strongly
anisotropic and vanishes in particular directions in the Brillouin zone. Moreover, the 
phase--sensitive superconducting interference device experiments \cite{squid}
demonstrated the sign change of the order parameter between the 
$x$ and $y$ directions. Generally, these results are consistent
with the $d_{x^2-y^2}$ pairing scenario. On the other hand there
are experimental indications, which had questioned the pure $d_{x^2-y^2}$ 
symmetry of the energy gap and suggest mixed pairing symmetry 
with a dominant $d$--wave component 
(e.g., $d\pm s$ or $d\pm is$) \cite{kouznetsov,krishan,sun,ma}. 

The measurement of the upper critical field can
give insight into the microscopic parameters 
of a relevant model. For example, the 
coherence length $\xi$ is usually derived indirectly from the expression 
$H_{c2}(0)=\phi_0/2\pi\xi^2$ \cite{hidaka}, where $H_{c2}(0)$ is the upper
critical field determined at $T=0$, and $\phi_0$ is the magnetic
flux quantum. The theoretical investigation of the upper critical
field for different pairing symmetries is predominantly based on the 
Ginzburg--Landau (GL) \cite{ginzburg} theory or the Lawrence-Doniach 
\cite{doniach}
approach in case of layered superconductors. With the help of linearized GL equations
Won and Maki\cite{Won} have shown that $H_{c2}$ in a model 
with repulsive on--site interaction depends linearly on
temperature near $T_c$ and saturates at $T\rightarrow 0$.
They have not found any sign of the upward behavior. There are also 
calculations for $H_{c2}$ in systems with mixed symmetries,
especially for superconductors in which the dominant $d$--wave
order parameter coexists with a subdominant $s$--wave component.
However, in the most of these approaches $H_{c2}(T)$ exhibits negative curvature.
On the other hand, results obtained in Ref. \cite{dai} suggest
that the upward curvature of the critical field could be a characteristic
feature of a $d$--wave superconductor. The positive curvature of $H_{c2}(T)$ can also
originate from the presence of magnetic impurities \cite{kresin1,kresin2}.   

A separate problem, that is usually neglected in the above approaches,
is the influence of the periodic lattice potential on the upper critical
field \cite{my}. Application of magnetic field to the two--dimensional (2D)
electron system in a tight--binding approximation  leads to a fractal 
energy spectrum known as Hofstadter's butterfly, where even very small changes
in magnetic field can result in a drastic changes of the spectrum 
\cite{harper,langbein,hofstadter}. In this paper we investigate 
the upper critical field for electrons described by the two--dimensional     
tight--binding model with intra-- and intersite pairing. We show that 
anisotropic superconductivity is less affected by the external magnetic field
then the isotropic one. We also demonstrate that the lattice effects
can give rise to important corrections with respect to 
Helfand--Werthamer \cite{hw,kresin1} solution of the Gor'kov equations \cite{gorkov}.
This effect is of particular importance in the vicinity of the 
van Hove singularity and at low temperatures.
 
\section{Gap equation close to $H_{c2}$}

We consider a two--dimensional square lattice immersed in a uniform,
perpendicular, magnetic field. The BCS--type Hamiltonian is of the form  
\begin{eqnarray}
\hat{H} &=& \hat{H}_{\rm kin} + \hat{H}_V
-\mu \sum_{i,\sigma} c^{\dagger}_{i\sigma}c_{i\sigma} \nonumber \\ 
&& -g\mu_{B}H_{z}\sum_{i} 
\left(c^{\dagger}_{i\uparrow}c_{i\uparrow}
-c^{\dagger}_{i\downarrow}c_{i\downarrow} \right),
\end{eqnarray}
where $c^{\dagger}_{i \sigma}$ ($c_{i \sigma}$) creates 
(annihilates) an electron with spin $\sigma$ on site $i$.
The chemical potential $\mu$ is introduced in order to
control the doping level. The last term in the above Hamiltonian
describes the paramagnetic Pauli coupling to the external field.
Here, $g$ stands for the gyromagnetic ratio, $\mu_{B}$ is
the Bohr magneton and $H_z$ is the $z$--component of the
external field. The first ($\hat{H}_{\rm kin}$) and the second 
($\hat{H}_V$) term in the Hamiltonian represents the kinetic energy
and the pairing interaction, respectively. Within the tight--binding
approach    
\begin{equation}
\hat{H}_{\rm kin} = \sum_{<ij>,\sigma} t_{ij}\left(\Av \right) 
c^{\dagger}_{i\sigma}c_{j\sigma}.  
\end{equation}
The electrons are gauge--invariantly coupled with local $U(1)$
gauge field by a phase--factor in
the kinetic--energy hopping term.  According
to the Peierls substitution
\cite{peierls} in the presence of magnetic field 
the original hopping integral between sites $i$ and $j$, $t_{ij}$
acquires an additional factor
\begin{equation}
t_{ij}\left(\Av \right)= t_{ij}
\exp\left(\frac{ie}{\hbar c} \int^{\Rv_{i}}_{\Rv_{j}}
\Av\cdot d\lv\right).
\end{equation} 
In the case of the on--site pairing,
which leads to isotropic order parameter, 
the BCS--type interaction takes on the form
\begin{equation}
\hat{H}_V=-\:
V\sum_{i}\left(c^{\dagger}_{i\uparrow}c^{\dagger}_{i\downarrow}\Delta_{i}
+c_{i\downarrow}c_{i\uparrow} \Delta^{\star}_{i} \right).
\end{equation}
Here, we have introduced local superconducting order parameter,
$\Delta_{i}=\langle c_{i\downarrow}c_{i\uparrow}\rangle$,
which in the presence of the magnetic field can 
change from site to site \cite{my}.  
We also consider anisotropic superconductivity
with the intersite pairing interaction given by
\begin{equation}
\hat{H}_V=-\:
V\sum_{<ij>}\left(c^{\dagger}_{i\uparrow}c^{\dagger}_{j\downarrow}
\Delta_{ij}
+c_{i\downarrow}c_{j\uparrow} \Delta^{\star}_{ij} \right).
\end{equation}
For the sake of simplicity we restrict our considerations
only to the nearest--neighbor coupling 
with  the singlet order parameter $\Delta_{ij}=\langle c_{i\downarrow}c_{j\uparrow}
-c_{i\uparrow}c_{j\downarrow}\rangle$.

We start with the discussion of the normal state properties. Similarly 
to Ref. \cite{my} we make use of an unitary transformation $U$ that 
diagonalizes the kinetic part of the Hamiltonian
\begin{equation}
U^\dagger \hat{H}_{\rm kin} U = {\cal H}_{\rm kin}.
\end{equation}
This transformation defines a new set of fermionic operators 
$a_{n\sigma}=\sum_i U^\dagger_{ni} c_{i\sigma}$, in which
the Hamiltonian in the normal state takes on the diagonal form
\begin{equation}
{\cal H} = \sum_{n\sigma} \left( E_{n} - \mu -\sigma g \mu_B H_z \right)
a^\dagger_{n\sigma} a_{n\sigma}.
\end{equation} 
In the absence of the magnetic field $U$ represents transformation from
the Wannier to the Bloch representation. 
For finite magnetic field and general gauge
the quantum number $n$ enumerates eigenstates,
although does not represent a reciprocal lattice vector. 
In order to simplify further discussion we
restrict our considerations only to the nearest neighbor hopping with
$t_{\langle ij \rangle} \equiv -t$. We also assume the type--II limit of 
superconductors where the magnetic field can be regarded as 
a spatially uniform object. 
Choosing the Landau gauge ${\bf A} = H_z(0,\,x,\, 0)$
the hopping integral depends explicitly only on $x$ and
the momentum in $y$ direction $p_y$ remains a good quantum number.
Due to the plane--wave behavior in $y$ direction
the unitary matrix $U$
takes on the form
\begin{equation}
U_{i\left(\bar{p}_{x},p_{y}\right)}=
U_{\left(x,y\right)\left(\bar{p}_{x},p_{y}\right)}=
N^{-1/4}\;e^{\displaystyle ip_{y}ya}
g\left(\bar{p}_{x},p_{y},x\right),
\end{equation}
where $(ax,ay)$ is the position of the $i$--th site and 
$\left(\bar{p}_x,p_y\right)$ represents $n$--th
eigenstate of the Hamiltonian (7). 
Straightforward calculations \cite{hofstadter} show that the
$x$--dependent part of the wave function $g\left(\bar{p}_{x},p_{y},x\right)$
fulfills a one--dimensional difference equation
\begin{eqnarray}
&& g\left(\bar{p}_{x},p_{y},x+1\right)+
2\cos\left(h x -p_{y}a\right)g\left(\bar{p}_{x},p_{y},x\right) \nonumber \\ 
&& + g\left(\bar{p}_{x},p_{y},x-1\right) 
=t^{-1}E\left(\bar{p}_{x},p_{y}\right)
g\left(\bar{p}_{x},p_{y},x\right),
\end{eqnarray}
where  we have introduced a reduced dimensionless 
magnetic field
$h=e a^{2}H_{z}/(\hbar c)$. This quantity can be expressed with the
help of magnetic flux $\phi$ through lattice cell and flux 
quantum ($h=2 \pi \phi/\phi_0$). Equation (9) is known as the Harper equation 
\cite{harper} and has extensively been studied \cite{harpeq,hasegawa}.
The Harper equation, derived here within a tight-binding approximation,
can be also obtained in a case of weak perturbation
of a Landau--quantized two--dimensional electron 
system\cite{langbein,thouless}.

Now, let us take into account the pairing potential $H_V$. In order to 
investigate the transition from the superconducting to the normal state
we make use of equation of motion for the anomalous Green function. In the
case of the isotropic on--site pairing one obtains 
\begin{eqnarray}
&&\left[\omega-E\left(\bar{p}_{x},p_{y}\right)
+\mu+g\mu_BH_z \right]
\langle\langle a_{\left(\bar{p}_{x},p_{y} \right) \uparrow}\mid 
a_{\left(\bar{k}_{x},k_{y}\right) \downarrow} \rangle\rangle \nonumber \\
&&= -V \sum_{i,\bar{k}^{\prime}_{x},k^{\prime}_{y} } \Delta_{i}
U^{\star}_{i\left(\bar{p}_{x},p_{y} \right)} 
U^{\star}_{i\left(\bar{k}^{\prime}_{x},k^{\prime}_{y}\right) } 
\langle\langle 
a^{\dagger}_{\left(\bar{k}^{\prime}_{x},k^{\prime}_{y}\right) \downarrow}\mid 
a_{\left(\bar{k}_{x},k_{y}\right) \downarrow} 
\rangle\rangle. \nonumber \\
\end{eqnarray}
As far as we are close to the phase transition we make use of a linearized gap 
equation i.e., we calculate the propagator $\langle\langle 
a^{\dagger}_{\left(\bar{k}^{\prime}_{x},k^{\prime}_{y}\right) \downarrow}\mid 
a_{\left(\bar{k}_{x},k_{y}\right) \downarrow} 
\rangle\rangle$ in the normal state. Similarly to the standard BCS theory,
such approach allows one to determine the critical temperature or, in 
our case, the upper critical field. However, it is irrelevant for calculations below 
$T_c$.

The choice of the Landau gauge implies that the isotropic order parameter
does not depend on $y$: $\Delta_i \equiv \Delta_{(x,y)} = \Delta_x$.
Then, the linearized gap equation reads
\begin{equation}
\vec{\Delta} = {\cal M}\vec{\Delta},
\end{equation}
where $\vec{\Delta} = (\Delta_1,\ \Delta_2,\ \Delta_3, ...)$ and
\begin{eqnarray}
{\cal M}(x,x') & = & \frac{V}{\sqrt{N}} \sum_{\bar{p}_x,p_y,\bar{k}_x} 
g(\bar{p}_x,p_y,x) \: g(\bar{k}_x,-p_y,x) \nonumber \\ 
&\times& g(\bar{k}_x,-p_y,x')\: g(\bar{p}_x,p_y,x') \: \chi(\bar{p}_x,p_y;
\bar{k}_x,-p_y). \nonumber \\ 
\end{eqnarray}
In the presence of the magnetic field the Cooper pair susceptibility 
is given by
\begin{eqnarray}
\chi(\bar{p}_x,p_y;\bar{k}_x,k_y)& =& \left[\tanh\frac{E\left(\bar{p}_{x},
p_{y}\right)-\mu-g\mu_BH_z}{2k_{B}T}
 \right. \nonumber \\
&&\left. +\tanh\frac{E\left(\bar{k}_{x},k_{y}
\right)-\mu+g\mu_BH_z }{2k_{B}T}
\right] \nonumber \\
&& \times
\left[2\left(E\left(\bar{p}_{x},p_{y}\right)+E\left(\bar{k}_{x},
k_{y}\right)-2\mu\right) \right]^{-1}. \nonumber \\
\end{eqnarray}

In the case of the nearest--neighbor pairing we obtain the gap equation 
analogous to Eq. (11). Similarly to the isotropic
pairing $\Delta_{ij}$ does not depend explicitly on $y$. However, there are
two types of order parameter at each site: $\Delta^{(x)}_{x}$ when sites $i$ and
$j$ lay along the $x$ axis, and  $\Delta^{(y)}_{x}$ when sites $i$ and
$j$ lay along the $y$ axis. Close to the upper critical field the gap equation
for anisotropic superconductivity can be written in a matrix form
\begin{equation}
\left(\begin{array}{c} \vec{\Delta}^{(x)} \\ \vec{\Delta}^{(y)} \end{array}
\right)=\left(\begin{array}{cc}{\cal M}^{\left(x,x\right)}&{\cal M}^{\left(x,y\right)}
\\ 
{\cal M}^{\left(y,x\right)}&{\cal M}^{\left(y,y\right)}\end{array}\right)
\left(\begin{array}{c} \vec{\Delta}^{(x)} \\ \vec{\Delta}^{(y)} \end{array}
\right),
\end{equation}
where 
\begin{eqnarray}
{\cal M}^{\left(\alpha,\beta\right)}(x,x^{\prime}) 
& = & \frac{V}{\sqrt{N}} \sum_{\bar{p}_x,p_y,\bar{k}_x}
\chi(\bar{p}_x,p_y;\bar{k}_x,-p_y) \nonumber \\
&& \times A^{\left(\alpha\right)}\left(\bar{p}_x,\bar{k}_x,p_y,x\right)
A^{\left(\beta\right)}\left(\bar{p}_x,\bar{k}_x,p_y,x^{\prime}\right), \nonumber \\
\end{eqnarray}
and
\begin{eqnarray}
A^{\left(x\right)}\left(\bar{p}_x,\bar{k}_x,p_y,x\right)&=&
  g(\bar{p}_x,p_y,x) g(\bar{k}_x,-p_y,x+1) \nonumber \\
 &+&  g(\bar{p}_x,p_y,x+1) g(\bar{k}_x,-p_y,x), \nonumber \\
 && \\
A^{\left(y\right)}\left(\bar{p}_x,\bar{k}_x,p_y,x\right)&=& 2 \cos( p_y a) \nonumber \\ 
&\times& g(\bar{p}_x,p_y,x) g(\bar{k}_x,-p_y,x).
\end{eqnarray}

Equations (11) and (14) constitute a system of linear equations for the order
parameters and the condition for existence of a non--zero solution
can be written as
\begin{equation}
\det \left({\cal M} - I\right) = 0
\end{equation}
in the case of isotropic pairing, and 
\begin{equation}
\det \left(\begin{array}{cc}{\cal M}^{\left(x,x\right)} - I&
{\cal M}^{\left(x,y\right)}\\
{\cal M}^{\left(y,x\right)}&{\cal M}^{\left(y,y\right)} - I\end{array}\right) = 0
\end{equation}
for anisotropic superconductivity, where $I$ is the unit 
matrix. These equations allow one to obtain the magnitude of the upper critical
field perpendicular to the plane. For the two--dimensional square lattice the size 
of matrices which enter Eqs. (18) and (19) is proportional to the square root of
the number of the lattice sites. Analytical
solutions of the Harper equation (9)
are known only in a few cases of commensurable field\cite{hasegawa} 
(in our notation $h=2\pi p/q$, where $p$ and $q$ are relative prime integers),
which correspond to unphysically high magnetic field. Therefore, in order to 
investigate $H_{c2}$
we restrict our considerations to a finite lattice, for which we are able to
analyze numerically the commensurable and incommensurable magnetic field on an
equal footing. 

\section{Discussion of results}
We consider square $M\times M$ cluster with periodic boundary conditions (bc)
along the $y$ axis. As the Landau gauge breaks the translation invariance along $x$ axis
we use fixed bc in this direction. 
An additional advantage originating from such a mixed bc is the absence of the
unphysical degeneracy of states at the Fermi level, which occurs for the
half-filled band in cluster calculations with fixed or periodic bc taken in both
directions \cite{jokojama}. In order to estimate the finite size effects 
we have carried out numerical calculations for clusters of different sizes.  
We have found that in the case of the isotropic pairing and
small concentration of holes ($\delta < 0.2$)  
there are no significant 
differences between results obtained on $150 \times 150$
and $200 \times 200$ clusters. For anisotropic pairing already $120 \times 120$ clusters
give convergent results.         
\begin{figure}[h]
\centerline{\epsffile{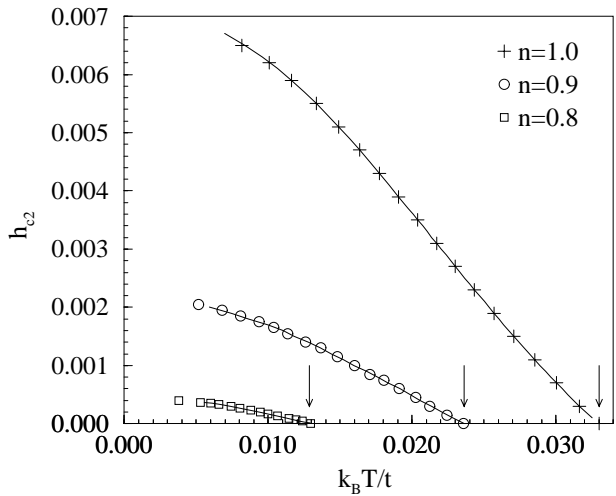}}
\caption{Temperature dependence of the reduced upper critical field for isotropic 
pairing and different occupation numbers $n$. The cross, circle and square marks
indicate results obtained on $150 \times 150$ cluster, whereas the solid lines 
correspond to $200 \times 200$ cluster. The arrows show the superconducting transition
temperature for an infinite system calculated from the BCS gap equation in the absence
of magnetic field. $V=t$ has been assumed.}
\end{figure}   

\begin{figure}[h]
\centerline{\epsffile{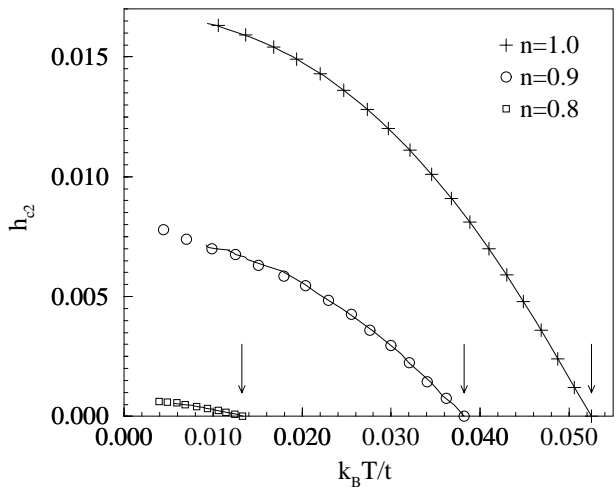}}
\caption{The same as in Fig. 1. but for anisotropic pairing. The cross, circle and 
square marks
indicate results obtained on $120 \times 120$ cluster, whereas the solid lines 
correspond to $150 \times 150$ cluster. The arrows show the $d$--wave superconducting 
transition temperature for an infinite system calculated from the BCS gap equation 
in the absence of magnetic field. Here, $V=0.3\,t$ has been assumed.}
\end{figure}   

Figures 1. and 2. show the reduced critical field, $h_{c2}=e a^{2}H_{c2}/(\hbar c)$,
for different concentrations of holes.  Independently on the symmetry of the 
superconducting order parameter i.e., for isotropic (Fig. 1.) as well as anisotropic 
 pairing (Fig. 2.), the slope of $H_{c2}(T)$ strongly decreases with 
increasing doping. 
Note that our cluster results exactly reproduce the BCS transition temperature
when the magnetic field tends to zero. In the case of intersite pairing
the arrows indicate BCS solutions for $d_{x^2-y^2}$ superconductivity. 
However, the external magnetic field affects the relative phases of the order
parameter in the $x$ and $y$ directions, which can change from site to site.
Therefore, it is impossible to determine globally the type of the symmetry of the
energy gap in the presence of magnetic field.

Contrary to the conclusion presented in Ref. \cite{dai},
our results (Figs. 1. and 2.)  do not indicate that
the upward curvature of $H_{c2}(T)$ can emerge as a direct
consequence of the symmetry of superconducting state.
However, the anisotropy of the order parameter can
significantly influence the magnitude of the upper critical field.
In order to investigate this relationship we have directly compared
results obtained for on-- and intersite pairing
for isotropic and anisotropic superconductivity.
We have chosen the magnitudes of the pairing potentials $V$,
which, in the absence of magnetic field,
lead to the same superconducting transition temperatures
for isotropic and anisotropic superconductivity.
Fig 3. shows the temperature dependence of the upper critical
field obtained for the half-filled case. One can see
that the anisotropic superconductivity is less
affected by the external field than the isotropic one.
\begin{figure}[h]
\centerline{\epsffile{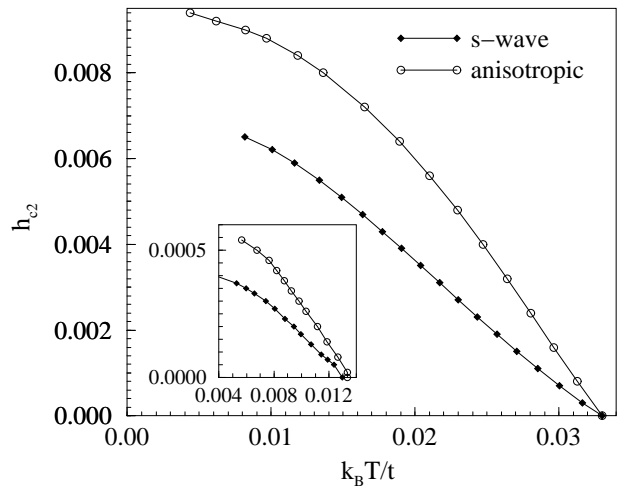}}
\caption{The temperature dependence of the upper critical
field evaluated for the half-filled case, $n=1$.
The circle and diamond symbols denote results obtained
for intersite pairing with $V=0.244t$ and
on--site pairing with $V=t$, respectively.
The symbols and continuous lines correspond to the
same sizes of clusters as in Figs. 1 and 2. The inset shows
the upper critical field obtained for the occupation number $n=0.8$.
Here, we have taken the intersite pairing potential $V=0.3t$.}
\end{figure}
\noindent
An important observation is that
this result depends neither on the magnitude of the pairing potential
nor on the concentration of holes (see the inset in Fig. 3). Therefore,
it can be considered as a characteristic feature of the
two--dimensional lattice gas.

In the absence of magnetic field there is a van Hove
singularity in the middle of the band. Although, the
external field results in a splitting
of the Bloch band into a huge number of subbands,
the presence of the  original van Hove singularity
is reflected in the Hofstadter spectrum \cite{hofstadter}.
In contradistinction to the structure  of Landau levels,
the Hofstadter spectrum does not consist of
uniformly distributed energy levels. In particular,
the average distance between the energy levels
 close to the Fermi energy achieves its
minimum when the chemical potential is in the
middle of the Bloch band. It can be considered
as a remnant of the original van Hove singularity.
\begin{figure}[h]
\centerline{\epsffile{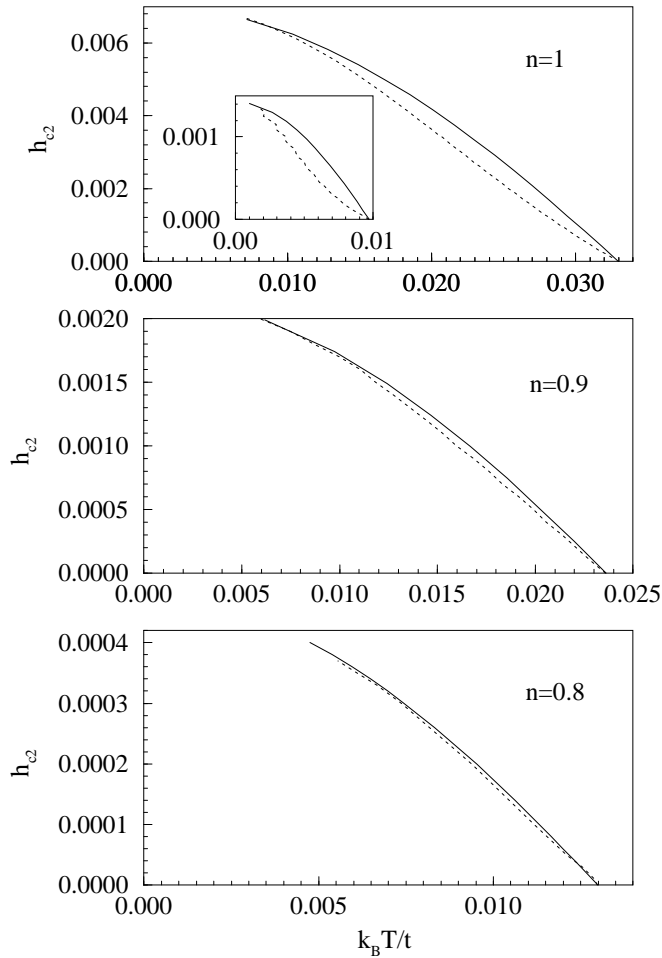}}
\caption{The upper critical field for the on--site pairing
(dashed lines)
fitted to the results obtained for the two--dimensional version of the
Helfand--Werthamer approach to the Gor'kov equations (continuous lines).
We have chosen the pairing potential $V=t$, for different values
of the occupation number $n$, indicated in the figures.
The inset shows results obtained for $V=0.7t$ and $n=1$.}
\end{figure}
\noindent
The question which arises concerns the impact of this
feature on the upper critical field. In order to
analyze this problem we have fitted $H_{c2}(T)$
obtained for isotropic superconductivity to the results obtained for
the two--dimensional version \cite{kresin1} of the
Helfand--Werthamer approach to the  Gor'kov equations.
Fig. 4 shows the numerical results.
Away from the half--filled case the qualitative temperature
dependence of the upper critical field can be very well approximated
by the solution of the Gor'kov equations.
It suggests, that the complicated Hofstadter spectrum
does not influence the temperature dependence of the critical
field, provided that the Fermi level is  far enough
from the original van Hove singularity. However, in the vicinity
of the van Hove singularity the second derivative of $H_{c2}(T)$
is significantly enhanced, when compared to the results obtained
from the Gor'kov equations. It is of particular importance
for small values of the pairing potential, when the system
remains in superconducting state only at relatively low temperatures
and the Cooper--pair susceptibility is strongly peaked at the
Fermi level. Then, the curvature of $H_{c2}(T)$ can gradually
change from negative to positive, as depicted in the inset in Fig. 4.
This effect takes place for isotropic as well as for anisotropic pairing.
Similar results have been reported in Ref. \cite{dias}.
\section{Concluding remarks}

We have investigated the temperature dependence of the
upper critical field for the two--dimensional lattice
gas. With the help of unitary transformation we have
obtained a diagonal form of the Hamiltonian in the normal
state and derived gap equations both for isotropic and
anisotropic superconductivity. We have discussed
influence of the symmetry of the superconducting state
and the van Hove singularity on the upper
critical field. Our results clearly indicate that
the symmetry of the superconducting order parameter
itself can not lead to upward curvature of $H_{c2}(T)$.
However, quite pronounced tendency can be observed
for the half-filled case, when the
Fermi energy is close to the original van Hove
singularity. In the absence of the external field
this singularity occurs in the middle of the band.
The enhancement of curvature of $H_{c2}(T)$
takes place for isotropic as well as anisotropic
superconductivity and is of particular importance for
small values of the pairing potential. Then, the
curvature can gradually change from negative to
positive. This effect smears out for larger doping where
the temperature dependence of the upper critical
field can be rendered very well when solving the Gor'kov
equations. We have found that in the case of anisotropic
pairing the upper critical field exceeds the critical field
obtained for isotropic superconductivity. It takes place
for small doping ($\delta < 0.2$) and arbitrary magnitude
of the pairing potential. These results suggest that in
the two--dimensional lattice gas anisotropic superconductivity
is less affected by the external field than the isotropic one.

The proposed method allows one to derive the gap equation
in the same way as the standard BCS approach. The only differences
are related to the fact that the diagonal form of the normal--state
Hamiltonian is obtained numerically and the superconducting
order parameter can be a site-dependent quantity. The similarity
between our method and the BCS approach
allows for straightforward incorporation of the local Coulomb repulsion
within any standard approximation.
Here, one may expect destructive influence of correlations, in
particular in the isotropic channel. This originates from the fact
that local repulsion always acts to the detriment of the formation
of local Cooper pairs. The impact of Coulomb, Hubbard--like,
correlations on anisotropic superconductivity seems to depend on
the approximation scheme. This problem is under our current investigation.

\acknowledgments
This work has been supported by the Polish State Committee
for Scientific Research. We acknowledge a fruitful discussion with
Janusz Zieli{\'n}ski.

\end{document}